\documentclass{ifacconf}

\usepackage{graphicx}     
\usepackage{natbib}       
\usepackage{amsmath}
\usepackage{chemmacros}
\usepackage{booktabs}
\usepackage{physics}
\usepackage{subcaption}
\usepackage{microtype}

\definecolor{mygreen}{rgb}{0.3,0.6608,0.5471}
\definecolor{myorange}{rgb}{0.9882,0.5529,0.3843}
\definecolor{myblue}{rgb}{0.3529,0.4275,0.8961}
\definecolor{mygrey}{rgb}{0.6, 0.6, 0.6}

\begin{document}
\def\subinrm#1{\sb{\mathrm{#1}}}
{\catcode`\_=13 \global\let_=\subinrm}
\mathcode`_="8000
\def\supinrm#1{\sp{\mathrm{#1}}}
{\catcode`\^=13 \global\let^=\supinrm}
\mathcode`^="8000
\def\upsubscripts{\catcode`\_=12 } \def\normalsubscripts{\catcode`\_=8 }
\def\upsupscripts{\catcode`\^=12 } \def\normalsupscripts{\catcode`\^=7 }
\upsubscripts
\upsupscripts

\renewcommand{\thefootnote}{} 

\begin{frontmatter}

\title{State estimation for gas purity monitoring and control in water electrolysis systems} 

\author{Lucas Cammann, Johannes Jäschke}
\address{Department of Chemical Engineering, Norwegian University of Science
and Technology Trondheim, NO-7491 (e-mail: lucas.cammann@ntnu.no,
johannes.jaschke@ntnu.no).}

\begin{abstract}              
Green hydrogen, produced via water electrolysis using renewable energy, is seen as a cornerstone of the energy transition. Coupling of renewable power supplies to water electrolysis processes is, however, challenging, as explosive gas mixtures (hydrogen in oxygen) might form at low loads. This has prompted research into gas purity control of such systems. While these attempts have shown to be successful in theoretical and practical studies, they are currently limited in that they only consider the gas purity at locations where composition measurements are available.  As these locations are generally positioned downstream of the disturbance origin, this incurs considerable delays and can lead to undetected critical conditions. In this work, we propose the use of an Extended Kalman Filter (EKF) in combination with a simple process model to estimate and control the gas composition at locations where measurements are not available. The model uses noise-driven states for the gas impurity and is hence agnostic towards any mechanistic disturbance model. We show in simulations that this simple approach performs well under various disturbance types and can reduce the time spent in potentially hazardous conditions by up to one order of magnitude.

\end{abstract}

\begin{keyword}
Water electrolysis, Hydrogen, HTO, State estimation
\end{keyword}

\end{frontmatter}

\section{Introduction}  
Green hydrogen is considered an attractive energy vector for the future energy transition. It is produced through electrolysis reactions consuming water and electrical energy from renewable sources. These renewable power sources can introduce considerable load fluctuations into the process, increasing the necessity for flexible operation. One major bottleneck to such flexible operation of water electrolysis processes is gas purity requirements that impose limits on the minimum load at which the process can be safely operated (\cite{brauns_alkaline_2020}).   \footnote{© 2025 the authors. This work has been accepted to IFAC for publication under a Creative Commons Licence CC-BY-NC-ND.}

\vspace{-0.075cm}
The gas purity of water electrolysis processes is commonly defined in terms of the oxygen-to-hydrogen ($OTH$) and the hydrogen-to-oxygen ratio ($HTO$), where often only the latter is used as it is typically higher. To maintain safety and operability of the process, the $HTO$ may not exceed 2\% in any part of the equipment, which is 50\% of the lower explosion limit for such gas mixtures. The $HTO$ typically rises when the load in the electrolysis stack is lowered. This is because the contamination mechanisms, e.g., diffusion and convection across the membrane, remain unchanged, while the \ch{O2} production is reduced. To extend the load range of water electrolysis systems, researchers have recently proposed different ways of controlling the $HTO$ in low load conditions. \cite{david_h2_2021} proposed an $\mathcal{H}_\infty$ optimal controller utilizing the separator pressure and liquid level to control the $HTO$. Use of pressure has been implemented on a \SI{0.5}{\cubic \metre \per \hour} experimental setup by \cite{qi_pressure_2021}, while a combined use of pressure and lye recirculation flowrate has been proposed by \cite{li_active_2022} and implemented on a \SI{4}{\kilo \watt} electrolysis system. They found that this combined approach reduces the minimum load from \num{20}\% of the nominal load to \num{8.95}\%. \cite{cammann_simple_2024} have proposed a constraint-switching control structure that utilizes both the pressure and lye flowrate to control the $HTO$, and \cite{hu_study_2024} similarly implemented a control strategy for both on the basis of an optimal operating curve. 

\vspace{-0.075cm}
The developed control methodologies show promise in increasing the flexibility of electrolysis processes; they  are, however, limited in that they rely on accurate measurements of the $HTO$. While measurement devices for the relevant binary gas mixtures are readily available and fast (response time $t_{90} =$ \SI{20}{\second} for a typical thermal conductivity binary gas transmitter), they require measurements to be taken in the pure gas phase. Consequently, compositions in a two-phase flow regime, as they occur, e.g., in the electrolysis stack or the piping, cannot be measured. This is especially critical as the most severe disturbances in the stack occur under these conditions. The first potential measurement location lies at the gas outlet of the gas-liquid separators, which can have a considerable gas inventory and therefore introduce additional delay to the measurement. Given the widely acknowledged importance of gas purity limits in water electrolysis, it is then surprising that, to the best of the authors knowledge, no work has been published concerning its state estimation. \\
In this work, we aim to close this research gap by developing a state estimator that enables monitoring and control of the $HTO$ in locations upstream of the separator. The developed state estimator is simple to implement and does not require a detailed model of the process or the nature of the disturbances. We show in simulations that this simple approach is effective in accurately estimating otherwise unmeasurable gas purity conditions. We further show that these composition estimates can be successfully used for inferential control, thereby reducing the time spent in potentially hazardous conditions by an order of magnitude when compared to the feedback case on the available measurement.

\begin{figure}
\includegraphics{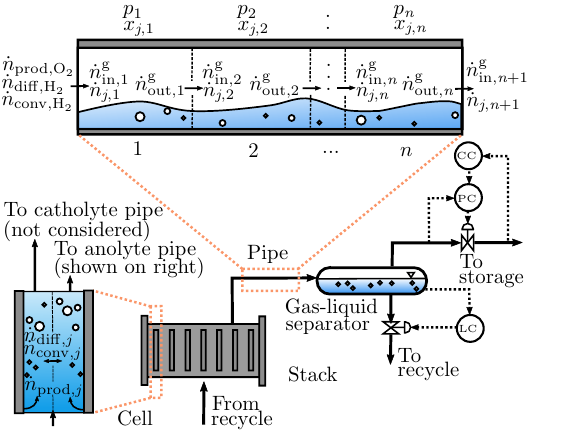}
 \caption{Sketch of the full plant model with magnified pipe and cell sections. 
 }
  \label{fig:System}
\end{figure}

\section{Model and methodology}
The following section presents the model of the system as well as the considered methodology for its estimation and control. This is preceded by a brief description of the considered water electrolysis system. While the developed methodology should be applicable to a broad range of electrolysis processes, we herein present it in the context of alkaline water electrolysis for the sake of concreteness.  
\subsection{System description}
The system under consideration is shown in Figure \ref{fig:System} and consists of an alkaline electrolyzer, a pipe, and a gas-liquid separator. Lye (30 \% aq. KOH) is fed to the electrolyzer, where water is consumed to create hydrogen (\ch{H2}) and oxygen (\ch{O2}) gas in the following overall reaction
\begin{align}
    \ch{H2O}(l) \ch{-> H2}(g) + \frac{1}{2} \ch{O2}(g).
\end{align}
Here, \ch{H2} is formed at the cathode side and \ch{O2} at the anode side of the cells. The anodic and cathodic effluents (anolyte and catholyte) are a mixture of product gases and lye, which have to be separated. This is done in gas-liquid separators that are typically designed with a horizontal aspect ratio and are sized to provide a residence time for the liquid lye of $\approx \SI{2}{\minute}$. Control of the liquid level $l$ and the pressure $p$ is classically done by manipulating the liquid outflow $\dot{m}_{lye}$, and the gaseous outflow $\dot{n}_{out}^{gas}$, respectively. In applications with a single electrolyzer stack, the separators are commonly placed on top of the stack, leading to a short length of the pipe. In large-scale systems, several stacks can be serviced by one set of separators, leading to increased pipe lengths and deadtimes in the system. 
To maintain safety and operability of the electrolysis process, the $HTO$ may not exceed 2\% in any piece of equipment. We herein refer to $HTO$ as the quotient of mole fractions $x$ or molar flowrates $\dot{n}$ in the gas phase between \ch{H2} and \ch{O2} (i.e., ${x_{\ch{H2}}}/{x_{\ch{O2}}}$ or $\dot{n}_{\ch{H2}}/\dot{n}_{\ch{O2}}$). Since composition measurements are only available under single-phase flow conditions, the composition of the gas stream exiting the separator is the first that can be measured after the electrolyzer stack. Here, disturbances that affect the composition are diffusion across the membrane as well as convection driven by pressure difference $\Delta p$. Other impurity mechanisms include the mixing of the anodic and cathodic lye streams to balance the concentration or out-of-norm conditions such as an undetected membrane fault or rupture in the electrolysis stack. Importantly, these contamination mechanisms are independent of the load of the stack itself. This means that these mechanisms become particularly critical at low current densities $I$, where less \ch{O2} is produced to sufficiently dilute the cross-contamination.
\begin{figure}
\centering
\includegraphics{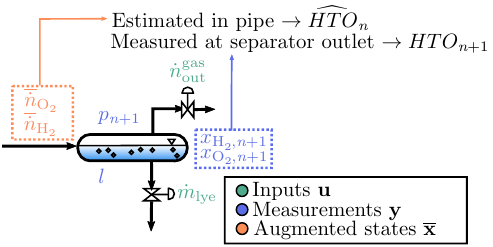}
\caption{Simplified model used for the state estimator.}
\label{fig:Estimation}
\end{figure}

\subsection{Full plant model}
The input disturbances affecting the composition are described by the relationships governing gas production, diffusion, and convection of species $j$ within the electrolysis stack. The former two are calculated as follows
\begin{align}
    \dot{n}_{prod,\textit{j}} &= \frac{IA_{c}}{z_{\textit{j}}F} \ , \ \ \,     \dot{n}_{diff,\textit{j}} = \frac{pS_{\textit{j}}D_{\textit{j}}A_{c}}{d_m} \quad \forall j \in \{\ch{H2,O2}\}, \label{Eq.diff} 
\end{align}
while the latter is calculated through the heuristic stating that the convective transport across the membrane grows proportionally to the pressure difference $\Delta p$ according to $\dot{n}_{conv,\textit{j}} \approx \dot{n}_{diff,\textit{j}}\Delta p/\SI{0.01}{\bar}$ (\cite{schalenbach_acidic_2016}). In Eq. \ref{Eq.diff}, $A_{c}$ is the active area of the electrodes, which is herein taken to be equal to the area available for the diffusion and assumed to be \SI{598}{\metre \squared} over all cells, while $d_m$ is the membrane thickness of \SI{5}{\milli \metre}. As the $HTO$ is of concern solely on the anodic side, diffusion and convection are only considered for the contaminant species of hydrogen. The diffusion coefficient $D_{\ch{H2}}$ is herein taken to be \SI{5.59e-9}{\metre \per \second \squared} (\cite{haug_process_2017}), and the solubility constants $S_{\ch{H2}}$ and $S_{\ch{O2}}$ are  \SI{8.84e-5}{\mol \per \kilo \gram \per \bar} and \SI{8.13e-5}{\mol \per \kilo \gram \per \bar}, respectively (\cite{knaster_solubility_nodate}).

Product and contaminant gases from the stack enter the pipe-separator system. This system is modeled as a distributed system, with $n$ pipe compartments of equal volume and one larger volume in the form of the separator (index $n+1$). For the sake of simplicity, we only model the liquid level in the separator tank and assume that there is a constant liquid level in the pipe. This gives rise to the following set of differential equations $\forall i \in \{1,...,n+1\}$
\begin{subequations}
\begin{align}
    p_{\textit{i}}' &= \frac{(\dot{n}_{in,\textit{i}}^g - \dot{n}_{out,\textit{i}}^g)RT - p_{\textit{i}} V_{\textit{i}}'}{V_{\textit{i}}}, \\ \label{State1}
    l ' &= \frac{\dot{m}_{in} - \dot{m}_{lye}}{\rho A}, \\
    x_{\ch{H2},\textit{i}}' &= \frac{RT(\dot{n}_{\ch{H2},\textit{i}}x_{\ch{O2},\textit{i}}-\dot{n}_{\ch{O2},\textit{i}}x_{\ch{H2},\textit{i}})}{p_{\textit{i}}V_{\textit{i}}} , \\ 
    x_{\ch{O2},\textit{i}}' &= -  x_{\ch{H2},\textit{i}}'.\label{State4}
\end{align}
\end{subequations}

Here, we assume a constant operating temperature $T$ of \SI{80}{\celsius}. Further, the pipe length $l_p$ is taken to be \SI{1}{\metre} of $n = 5$ equal segments with the gas-liquid separator assumed to have a total volume $V_t$ of \SI{2}{\cubic \metre}. The separator is assumed to be rectangular, for which it generally holds that 
\begin{align}
    V_t = V + lA,
\end{align}
where the gas volume $V$ nominally occupies half of $V_t$, and $A$ is the horizontal area of the equipment. The gaseous outflow of the separator is manipulated by a pressure controller, whereas the flow through the preceding segments is governed by the pressure difference according to the Hagen-Pouiseuille equation for gaseous, single-phase flow
\begin{align}
   \dot{n}^g_{out,\textit{i}} =  {\frac {n\pi r^{4}}{16 \eta  R Tl_p}}\left({p_{\textit{i}}}^{2}-{p_{\textit{i}+1}}^{2}\right) \quad \forall i \in \{1,...,n\}.
\end{align}
Lastly, conservation of mass in each balance volume is ensured through the following relations
\begin{subequations}    
\begin{align}
    \dot{n}_{in,1}^g &= \dot{n}_{prod,\ch{O2}} + \dot{n}_{diff,\ch{H2}} + \dot{n}_{conv,\ch{H2}}, \\
     &= \dot{n}_{\ch{O2},1} + \dot{n}_{\ch{H2},1}, \nonumber\\
        \dot{n}_{out,\textit{i}}^g &=  \dot{n}_{in,\textit{i}+1}^g \, \, \, \, \,  \quad \forall i \in \{1,...,n\}, \\
     \dot{n}_{\textit{j},\textit{i}+1} &=  \dot{n}_{out,\textit{i}}^gx_{\textit{j,i}} \quad \forall i \in \{1,...,n\} , j \in \{\ch{H2},\ch{O2}\}, \\
    \dot{n}_{out,\textit{n}+1}^g &=  \dot{n}_{out}^{gas}, \\
         \dot{n}_{\ch{O2},\textit{n}+1} &=  \dot{n}_{\ch{O2},\textit{n}} - \underbrace{\dot{m}_{lye} S_{\ch{O2}}p_{\textit{n}+1}}_{\dot{n}_d} \label{Eq.SO2}.
\end{align}
\end{subequations}
We assume only oxygen to be dissolved in the lye in the gas-liquid separator, shown as $\dot{n}_d$ in Eq. \ref{Eq.SO2}. 
\subsection{State estimator and simplified model}
The state estimator is designed to be, \emph{a)}, simple to implement and, \emph{b)}, agnostic to any mechanistic description or potential measurement of disturbances $\mathbf{d}$. To do so, the full plant model is simplified by omitting equations upstream of the separator. Incoming flows of component gases are treated as augmented states $\overline{x}$ driven by process noise ($\mathbf{\overline{x}}_{\textit{k}} = \mathbf{\overline{x}}_{\textit{k}-1} + \mathbf{w}_{\textit{k}-1}$), giving rise to the following set of states $\mathbf{x}$, inputs $\mathbf{u}$ and measurements $\mathbf{y}$
\begin{subequations}
    \begin{align}
    \mathbf{x} &= [p_{\textit{n}+1}, \ l, \ x_{\ch{H2},\textit{n}+1}, \  x_{\ch{O2},\textit{n}+1}, \ \overline{\dot{n}}_{\ch{H2}}, \ \overline{\dot{n}}_{\ch{O2}}]^T,\\
               & = [p_{\textit{n}+1}, \ l, \ x_{\ch{H2},\textit{n}+1}, \ x_{\ch{O2},\textit{n}+1}, \ \overline{\mathbf{x}}]^T, \nonumber\\
    \mathbf{u} &= [\dot{n}_{out}^{gas}, \ \dot{m}_{lye}]^T, \\
    \mathbf{y} &= [p_{\textit{n}+1}, \ l, \ x_{\ch{H2},\textit{n}+1}, \ x_{\ch{O2},\textit{n}+1}]^T.
\end{align}
\end{subequations}
Figure \ref{fig:Estimation} shows a sketch of the system described by the resulting simplified model. This model is governed by six differential equations, those being Eq. \ref{State1} to \ref{State4} at index $i = n+1$ together with those associated with the augmented states. The state estimator of choice in this paper is the Extended Kalman Filter (EKF), which is chosen due to its simplicity and successful application in practice. The system model employed for the discrete EKF is defined as follows
\begin{align}
    \mathbf{x}_{\textit{k}} &= f(\mathbf{x}_{\textit{k}-1}, \mathbf{u}_{\textit{k}-1}) + \mathbf{w}_{\textit{k}-1} \quad \mathbf{w}_{\textit{k}} \sim \mathcal{N}(0,\mathbf{Q}), \\
        \mathbf{y}_{\textit{k}} &= h(\mathbf{x}_{\textit{k}}) + \mathbf{v}_{\textit{k}} \quad \mathbf{v}_{\textit{k}} \sim \mathcal{N}(0,\mathbf{R}). 
\end{align}
In this formulation $\mathbf{w}_{\textit{k}}$ and $\mathbf{v}_{\textit{k}}$ are the additive process and observation noise at timestep $k$, respectively, which are assumed to be multivariate Gaussian with zero mean and covariance matrices $\mathbf{Q}$ and $\mathbf{R}$. The state transition model $f$ is constructed by applying Euler discretization (time step $t_s$ = \SI{0.1}{\second})  to the simplified model, while the measurement model $h$ directly maps the measured states to their output channels. The prediction step of the EKF algorithm entails the following 
\begin{align}
    \mathbf{\hat{x}}_{\textit{k}} &= f(\mathbf{x}_{\textit{k}-1}, \mathbf{u}_{\textit{k}-1}), \\
    \mathbf{P}_{\textit{k}} &= F_{\textit{k}-1}\mathbf{P}_{\textit{k}-1}F_{\textit{k}-1}^T + \mathbf{Q}. \label{Eq.PUpdate}
\end{align}
Both the predicted state estimate $\mathbf{\hat{x}}_{\textit{k}}$ and the predicted covariance estimate $\mathbf{P}_{\textit{k}}$ are then updated according to 
\begin{align}
     \mathbf{\hat{x}}_{\textit{k}} &=  \mathbf{\hat{x}}_{\textit{k}-1} + \mathbf{K}_{\textit{k}}(\mathbf{y}_{\textit{k}}-h(\mathbf{x}_{\textit{k}})), \\
      \mathbf{P}_{\textit{k}} &= (\mathbf{I} - \mathbf{K}_{\textit{k}}H_{\textit{k}})\mathbf{P}_{\textit{k}},
\end{align}
where the Kalman gain $\mathbf{K}_{\textit{k}}$ is calculated as (\cite{doi:https://doi.org/10.1002/0470045345.ch13})
\begin{align}
    \mathbf{K}_{\textit{k}} = \mathbf{P}_{\textit{k}}H_{\textit{k}}^T(H_{\textit{k}}\mathbf{P}_{\textit{k}}H_{\textit{k}}^T+\mathbf{R})^{-1}. \label{Eq.KalmanGain}
\end{align}
Note that the terms $F$ and $H$ in Eqs. \ref{Eq.PUpdate} and \ref{Eq.KalmanGain} are defined as the Jacobians of the state transition and measurement model, giving rise to the following discrete state-space representation
\begin{align}
    \mathbf{x}_{\textit{k}} = \underbrace{F_{\textit{k}}}_{\text{\makebox[0pt]{$ \displaystyle \eval{\dv{f}{x}}_{\overset{\hat{\mathbf{x}}_{\textit{k}-1}}{\scriptscriptstyle{\hat{\mathbf{u}}_{\textit{k}-1}}}}$}}} \mathbf{x}_{\textit{k}-1} + \underbrace{G_{\textit{k}}}_{\text{\makebox[0pt]{$ \displaystyle \eval{\dv{f}{u}}_{\overset{\hat{\mathbf{x}}_{\textit{k}-1}}{\scriptscriptstyle{\hat{\mathbf{u}}_{\textit{k}-1}}}}$}}}\mathbf{u}_{\textit{k}-1} , \quad
    \mathbf{y}_{\textit{k}} = \underbrace{H_{\textit{k}}}_{\text{\makebox[0pt]{$ \displaystyle \eval{\dv{h}{x}}_{\scriptscriptstyle{\hat{\mathbf{x}_{\textit{k}}}}}$}}}\mathbf{x}_{\textit{k}}.
\end{align}
The respective matrices are found to be
\begin{flalign}
F  &= \left[\begin{smallmatrix}\frac{1}{t_s}- \frac{V'}{V} && \frac{Ap'}{V} && 0 && 0 && \frac{RT}{V} & \frac{RT}{V} \\
  0 && \frac{1}{t_s} &&0 && 0 && 0 & 0 \\
  \frac{x_{\ch{H2}}'}{p} && \frac{Ax_{\ch{H2}}'}{V} && \frac{1}{t_s} - a (\overline{\dot{n}}_{\ch{O2}}-\dot{n}_d) && a \overline{\dot{n}}_{\ch{H2}} && ax_{\ch{O2}} & -ax_{\ch{H2}} \\
    \frac{x_{\ch{O2}}'}{p}  && \frac{Ax_{\ch{O2}}'}{V} && a (\overline{\dot{n}}_{\ch{O2}}-\dot{n}_d) && \frac{1}{t_s} - a \overline{\dot{n}}_{\ch{H2}} && -ax_{\ch{O2}} & ax_{\ch{H2}} \\
  0 && 0 && 0 && 0 && \frac{1}{t_s} & 0 \\
  0 && 0 && 0 && 0 && 0 & \frac{1}{t_s}
  \end{smallmatrix}\right]t_s, \nonumber \\
G &= \left[\begin{smallmatrix}-\frac{RT}{V}&& \frac{p}{\rho V}\\
            0 && -\frac{1}{\rho A} \\
            0 && \frac{x_{\ch{H_2}}RTS_{\ch{O2}}}{V} \\
            0 && -\frac{x_{\ch{H2}}RTS_{\ch{O2}}}{V} \\
            0 && 0 \\
            0 && 0\end{smallmatrix}\right]t_s, \quad H = \left[\begin{smallmatrix}
            1&& 0 && 0 && 0 && 0 &&0\\
            0 && 1 && 0 && 0 && 0 &&0\\
            0 && 0 && 1 && 0 && 0 &&0 \\
            0 && 0 && 0 && 1 && 0 && 0\end{smallmatrix}\right].
\end{flalign}

For better readability, indexing subscripts are omitted above and the variable $a = RT/pV$ introduced. 
It can then be verified that the observability matrix 
\begin{align}
    \mathcal {O}=[H, \ \  HF, \ \ \dotsm, \ \ HF^{5}]^T
\end{align}
has full rank, which means the linear system is fully observable. For the simplified model, we assume the process and observation noise covariance matrices are time-invariant, where $\mathbf{R}$ is the identity matrix and 
\begin{align}
    \mathbf{Q} = \texttt{diag}\left(\num{10},\num{1},\num{e-4},\num{e-4},0.3,300\right).
\end{align}
The EKF is initialized with the steady-state values at nominal conditions and with the identity matrix as initial covariance estimate.

\subsection{Control structure and numerical implementation}
\begin{figure}
    \centering
\includegraphics{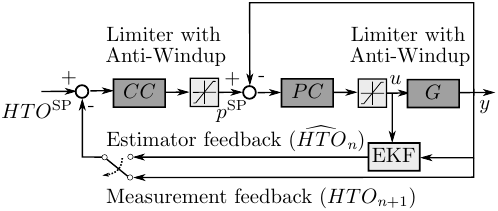}    
\caption{Block diagram of the concentration control structure. The logic switch indicates that either state, or measurement feedback is taken into account.}
    \label{fig:Feedback}
\end{figure}

The control structure used in this work is adapted from \cite{cammann_simple_2024}, who suggested a constraint-switching control structure in which both the pressure and the lye recirculation rate are manipulated to control the $HTO$ in the anodic gas-liquid separator. For the sake of simplicity in the presentation, we here consider only the pressure as a manipulated variable. This allows us to more clearly illustrate the influence of the different feedback paths while retaining an effective and simple control structure. 
The principal feedback loop consists of a cascade control structure in which the master concentration controller $CC$ updates the setpoint $p^{SP}$ of the slave pressure controller $PC$. This is shown in Figure \ref{fig:System} in the system sketch and in Figure \ref{fig:Feedback} as a block diagram. The concentration setpoint $HTO^{SP}$ is generally higher than the $HTO$ under nominal operating conditions but lower than the alarm limit of 2\%. In this work, it is set to 1.25\%. As the concentration controller will aim to increase the pressure when the measured gas purity is below its setpoint, a limiter is used to enforce sensible bounds on $p^{SP}$ based on limits of the downstream compression equipment. This configuration ensures that the pressure is normally high under nominal conditions, which is desired to reduce compression costs. While the idea to actively control the $HTO$ is relatively new, the few published works use direct feedback on the measurement at the gas outlet of the separator (here, $HTO_{\textit{n}+1}$). In this work, two different control configurations are compared: one employing such conventional measurement feedback and one using inferential control on an estimate of the gas purity state in the pipe ($\widehat{HTO}_{\textit{n}}$). The only difference between these structures is the state estimator in the measurement path, selected through a logic switch shown in Figure \ref{fig:Feedback}. 

All controllers are implemented as PI controllers using SIMC tuning rules (\cite{vilanova_simc_2012}). Tight tuning ($\tau_c = \theta$) has been applied to $PC$, with the closed-loop time constant of the outer loop being set to 15 times that of the inner loop. To apply the same tuning in both cases, measurement noise is only considered for the estimator feedback scenario. All simulations are carried out in \texttt{Matlab} using ode15s and default tolerances. 
   
\section{Simulation results}
We first show simulation results with the concentration controller in open-loop to illustrate how different disturbance types affect the system. The same disturbance sequence is then applied to the system in closed-loop, comparing the performance using feedback on either the gas purity measured after the separator or estimated in the pipe.
\subsection{Open-loop simulation}
\label{Sec: open-loop}
\begin{figure}
    \centering
     \includegraphics{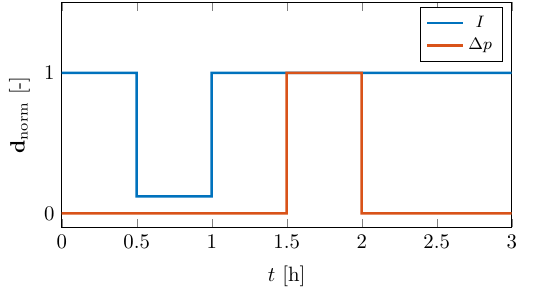}
     \includegraphics{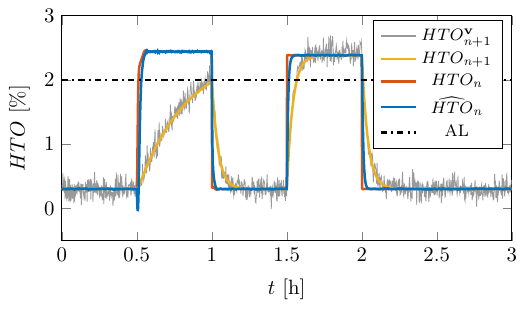}
          \caption{Upper panel: Normalized disturbance sequence. Lower pannel:
          True gas purity values in the pipe ($HTO_{\textit{n}}$) and separator ($HTO_{\textit{n}+1}$), and their estimate ($\widehat{HTO_{\textit{n}}}$) and noisy measurement ($HTO_{\textit{n}+1}^{\mathbf{v}}$), respectively,  together with the alarm limit (AL).} 
     \label{fig:OL-Sequence}
\end{figure}

The system is subjected to the disturbance sequence shown in the upper panel of Figure \ref{fig:OL-Sequence} with the concentration controller in open-loop ($p^{SP} = \SI{20}{\bar}$). To present different disturbance types together, the disturbances are normalized to their maximal value. The magnitude of the disturbances is chosen such that the $HTO$ exceeds the alarm limit (AL) of 2\% at steady-state. Following nominal operation, the current density $I$ is reduced at $t = \SI{30}{\minute}$. Such a condition could arise due to a scheduled load reduction or due to load variability when the system is directly coupled to a fluctuating power source. After nominal operation is resumed at $t = \SI{60}{\minute}$, a second disturbance enters the system at $t = \SI{90}{\minute}$. The increase of $\Delta p$ leads to increased convective transport across the membrane, a situation similar to a rupture of a membrane within the stack. After further $\SI{30}{\minute}$, this disturbance signal is also removed. 
\begin{figure*}[!ht]
    \centering
    \begin{subfigure}[t]{0.49\textwidth}
\includegraphics{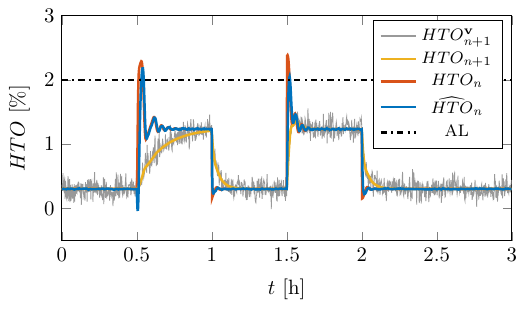}
\includegraphics{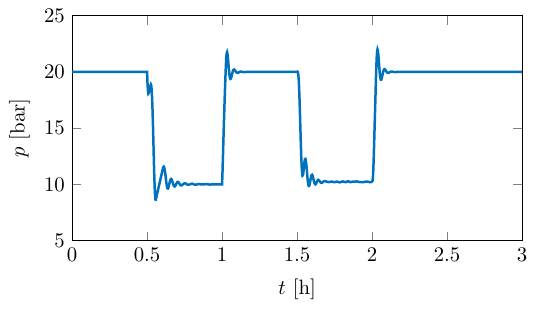}
    \caption{State feedback scenario}
    \label{fig:CL-ELF}
    \end{subfigure}
\hfill
    \begin{subfigure}[t]{0.49\textwidth}
       \includegraphics{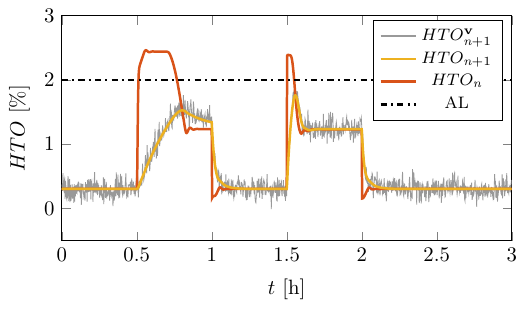}
       \includegraphics{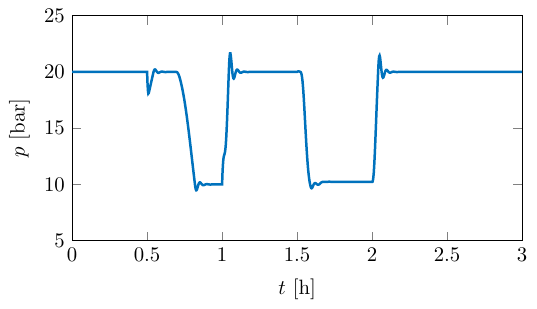}
    \caption{Measurement feedback scenario}
    \label{fig:CL}
    \end{subfigure}
    \caption{Simulation results for, (a), the state feedback scenario, and (b), the measurement feedback scenario. Upper panels:  True gas purity values in the pipe ($HTO_{\textit{n}}$) and separator ($HTO_{\textit{n}+1}$), and their respective estimate ($\widehat{HTO_{\textit{n}}}$) and noisy measurement ($HTO_{\textit{n}+1}^{\mathbf{v}}$)  together with the alarm limit (AL). Lower panels: Pressure as manipulated variable.}
    \label{fig:CLs}
\end{figure*}
The lower panel of Figure \ref{fig:OL-Sequence} shows the noisy gas purity measurement at the separator outlet ($HTO_{\textit{n}+1}^{\mathbf{v}}$), the ground truth as computed by the detailed process model at the measurement location ($HTO_{\textit{n}+1}$) and the last pipe segment ($HTO_{\textit{n}}$), as well as its estimate using the EKF ($\widehat{HTO}_{\textit{n}}$). Both disturbances lead to an increase in the $HTO$ above the alarm limit. In the first case, this increase is near instantaneous in the pipe connecting the stack to the gas-liquid separator, whereas it takes several minutes for the concentration measured at the outlet of the separator to approach a critical value. This can be attributed to the fact that a reduction in the current density increases the gas impurity and simultaneously reduces the gas production rate, slowing down the dilution in the separator. Conversely, the second disturbance leads to a more rapid increase of the $HTO$ in the separator, as here the gas production rate is unaffected. Importantly, the EKF is able to accurately estimate the unmeasurable gas concentration in both scenarios, i.e., $\widehat{HTO}_{\textit{n}}$ closely follows $HTO_{\textit{n}}$. In particular for the first case, this means that out-of-bound conditions can be detected minutes before they are measured and appropriate control action taken. Two things are further interesting to note about the presented profiles.  Firstly, the estimate of the augmented gas purity state appears slightly more noisy in the first disturbance case. This can be attributed to the chosen values for the constant process noise covariances, which are expected to perform differently for different gas flowrates. Secondly, the estimate of the augmented state shows an inverse response at the start of the first disturbance. Both will be shown to not affect the control performance in the next subsection, where simulation results are presented and discussed with the concentration control loop closed.

\subsection{Closed-loop simulation}
Figure \ref{fig:CLs} shows closed-loop simulation results for the disturbance shown in Subsection \ref{Sec: open-loop} considering feedback on the estimated gas purity in the pipe (Figure \ref{fig:CL-ELF}) and on the available $HTO$ measurement (Figure \ref{fig:CL}). In both subfigures the upper panel shows the relevant $HTO$ measurements and state estimates, while the lower panel shows the pressure used as a manipulated variable. 

Compared to the concentration controller being in open-loop, both feedback cases manage to reduce the time in which the gas purity in any process equipment exceeds the alarm limit. In the measurement feedback case, this is most prominent for the second disturbance after $t = \SI{90}{\minute}$ where the measurement is more rapidly affected by the upstream disturbance. During the first disturbance, however, the $HTO$ in the pipe remains prohibitively high for $\approx$ \SI{15}{\minute} as it is unmeasured and exhibits faster dynamics than in the separator. 
On the other hand, both disturbances are rapidly rejected in the state estimate feedback case. As in the open-loop scenario, the EKF predictions closely match the gas concentration in the pipe. Only at the initial peaks following the onset of each disturbance does the EKF underestimate the gas impurity. 
\begin{table}[!h]
\caption{Time out-of-bound ($t_{OOB}$) for open- and closed-loop simulations. Disturbances are differentiated by superscripts 1 and 2.}
\label{Tab:Res}
    \begin{tabular}{c c c c}
    \toprule 
    & Open-loop & $\widehat{HTO}_{\textit{n}}$ feedback & $HTO_{\textit{n}+1}$ feedback \\
    \midrule
    $t_{OOB}^1$ [min] & 29.3 &1.8 & 15.3\\
    $t_{OOB}^2$ [min] & 30 & 1 & 2.8\\
    \bottomrule
    \end{tabular}
\end{table}

\subsection{Quantitative comparison and discussion}
To more quantitatively assess the benefits of the proposed approach in increasing the process safety, we propose a performance indicator in the form of the time out-of-bound ($t_{OOB}$). We herein define $t_{OOB}$ as the time in which the $HTO$ in any part of the process exceeds the alarm limit of 2\%. This can be graphically interpreted as the time in which any of the ground-truth $HTO$ curves lie above the dotted alarm limit line in Figures \ref{fig:OL-Sequence} and \ref{fig:CLs}. The $t_{OOB}$ for both disturbance scenarios in the open-loop and the closed-loop cases is summarized in Table \ref{Tab:Res}. Unsurprisingly, the $t_{OOB}$ in the open-loop cases span effectively the entirety of the disturbance duration, as the disturbance in the stack nearly instantaneously propagates into the pipe. As shown previously, applying feedback control to the estimated gas purity upstream of the separator as opposed to its downstream measurement can greatly reduce the $t_{OOB}$. For the first disturbance, this reduction amounts to a factor of roughly 10, from \SI{15.3}{\minute} to \SI{1.8}{\minute}. As the second disturbance reaches the measurement location faster, the improvement potential is lower here but still considerable, with \SI{1}{\minute} compared to \SI{2.8}{\minute}.

Based on the simulation results, state estimation for gas purity in water electrolysis can, \emph{a)}, aid in detecting out-of-bound conditions where measurements are not available, and,  \emph{b)}, be used for inferential control of unmeasured states. As this greatly reduces the time spent in hazardous operating conditions, it should be considered in future control studies, as well as practical applications in the field of water electrolysis. To further investigate the potential of gas purity estimation in water electrolysis, further work is recommended to expand on the herein developed methodology. Firstly, this entails experimentally validating the proposed model and control approach. While the model mismatch between the full model and its simplified form was well handled by the EKF, a physical system might introduce more challenging and time-varying process and measurement noise characteristics. As the model is developed to be agnostic towards the disturbance type, it should similarly be investigated whether the proposed approach can be successfully applied to, e.g., polymer-electrolyte membrane electrolysis systems. To further reduce the $t_{OOB}$, alternative control approaches such as model predictive control should be investigated.
Lastly, it would be insightful to study the economic value of the proposed approach. Improvements in $HTO$ control would generally allow reducing the backoff towards the alarm limit, enabling more flexible operation at lower loads. This could therefore increase the online time of the system when considering direct input of renewable power or reduce the operating cost when participating in the day-ahead electricity market.

\section{Conclusions}
This work presents how state estimation can be used to improve the process safety in water electrolysis processes. Here, an Extended Kalman Filter (EKF) is used to estimate potentially dangerous gas impurities at locations where measurements are not available. The EKF uses a simplified process model and noise-driven states to capture essential impurity dynamics without relying on mechanistic disturbance models. Applying feedback to the estimated impurity state is shown in simulation to drastically reduce the time spent in potentially hazardous operating regions when compared to the standard implementation of measurement feedback. Depending on the disturbance type, this reduction can be up to nearly one order of magnitude, e.g., from \SI{15.3}{\minute} to \SI{1.8}{\minute}.

\begin{ack}
The authors would like to thank Vidar Alstad and Anushka Perera for fruitful discussions.  
\end{ack}

\normalsize
\bibliography{ifacconf}             
\end{document}